# Current Status of the Proposals for Unification of Notations and Guidelines for Data Presentation in the EMR Area – Blueprint for Future Actions

Czeslaw RUDOWICZ

Department of Physics and Materials Science, City University of Hong Kong, Hong Kong SAR

**Abstract.** The tremendous development of the electron magnetic resonance (EMR), i.e. EPR, ESR, and related spectroscopic techniques, and their applications in a number of fields is accompanied by a rather messy situation in the underlying theoretical framework. This is especially true concerning the EPR of transition ions in crystals, where an abundance of notations and lack of widely accepted clear definitions of the basic theoretical concepts characterize the current situation.  As a consequence, proliferation of a number of terminological misconceptions and incorrect relations for the zero-field splitting (ZFS) parameters has taken place.  Major problems detrimental for the future of EMR as well as the efforts aimed at alleviating these problems are reviewed. Rationales for coordinated activities in three major directions are outlined.  This includes (a) unification of spin Hamiltonian notations, (b) unification of EMR nomenclature, and (c) making better use of the various categories of data generated during EMR studies. Appropriate actions are proposed in each area, including ways of achieving internationally accepted notation standards, working out a glossary of EMR terms, setting up guidelines for presentation of EMR data, and establishment of EMR-related databases. Successful implementation of the proposals put forward here requires coordination of the activities at the international level in cooperation with the whole EMR community. Possible organizational framework for such activities and their coordination is also discussed.

## 1. Introduction

This paper is an extended and updated version of the Plenary Talk presented at the 31[th] Congress Ampere [1]. Improvements arising from the discussions held during the Congress have been incorporated into the paper. The problems plaguing some sub-areas of the electron magnetic resonance (EMR) spectroscopy, especially EPR of transition ions in crystals, are outlined.  Note that the term 'EMR' is used here to encompass the electron paramagnetic resonance (EPR) and electron spin resonance (ESR) as well as the related techniques, i.e. logically: EMR = EPR & ESR & more [2]. The up-to-now efforts and suggestions aimed at alleviating the problems in question are reviewed and appropriate actions are proposed. To provide a proper background for the proposals, an overview of the current situation in the EMR area is given in Section 2. In Section 3 and 4 we present in a nutshell the rationales for unification of spin Hamiltonian (SH) notations and unification of nomenclature used in the EMR area, respectively. These rationales stem from the need for rectification of the notational and conceptual problems outlined. In order to avoid an impression of personal criticism we concentrate on the nature of these problems, while referring the reader to the pertinent reviews or articles for source references. The various categories of data generated during EMR studies, which constitute 'community resources', are discussed in Section 5. Better utilization of these resources is a rationale for establishment of pertinent EMR-related databases. The proposals arising from the rationales presented in Sections 3 to 5 are summarized in Section 6. This includes (i) ways of achieving internationally accepted SH notation standards, (ii) working out a glossary of EMR terms, and (iii) setting up guidelines for data presentation in the EMR area.

These proposals are not only aimed at alleviating the notational and conceptual problems. Their successful implementation is an indispensable prerequisite for the fourth proposal, i.e. (iv) establishment of a computerized database of EMR data. Other related proposals are also discussed in Section 6. Efficient implementation of the proposals put forward in Section 6 requires coordination of the activities at the international level in cooperation with the whole EMR community. Focus activities for the proposed EMR support unit and its possible organizational framework are outlined in Section 6, thus forming a blueprint for future actions.

## 2. Overview of the current situation in the EMR area

A comparison of the articles collected in the Chapter "Historical Introduction" in [3] with those in the subsequent chapters in [3], as well as perusal of the contents of the Proceedings of recent EMR-related conferences, e.g. the Asia-Pacific EPR (ESR) Society's [APES] Symposia: APES' 97 [4], APES' 99 [5], and APES' 01 [6], provide a compelling evidence of the tremendous development of the experimental EMR techniques since the first observation of the 'paramagnetic resonance' spectrum by Zavoisky in 1944 (see, e.g. [7]). The resonance phenomenon consisted in transitions induced by an oscillating source of electromagnetic radiation ($\mathbf{B_1}$) between the energy levels of a system of electronic magnetic dipoles, described equivalently as a *'spin'* system, split by a static magnetic filed ($\mathbf{B_0}$). The perspective provided by the resources, e.g. in [3-7], shows the present depth and breadth of the applications of EMR spectroscopy. The *'cactus'* in Fig. 1 graphically represents the progression and growth of applications of EMR, from an experimental method originally invented in physics, to an important tool in various other fields of science. It appears, e.g. from [3-6], that the most robust leafs of the EMR (EPR/ESR) *cactus* belong at present to chemistry and materials science, and increasingly so to bio-medical applications of EMR. The growth of the EMR spectroscopy depicted in Fig. 1 was due to, as well as, was accompanied by, continuous advancement of various EMR-based experimental techniques driven by the expanding research opportunities. Various specialized monographs (for references, see [2]) may be consulted for specific aspects of modern applications of EMR evidenced in, e.g. [3-6].

On the theoretical side, the concept of spin Hamiltonian (SH) has become the cornerstone underlying all EMR sub-areas. Most sophisticated forms of SH are required to describe EMR of transition ions in crystals, see, e.g. [8-11], for which SH is anchored in the crystal (ligand) field (CF/LF) theory, see, e.g. [11-14]. The SH concept enables to quantify the experimental EMR spectra and parameterize them in terms of a set of SH parameters appropriate for a given 'spin' system. A critical review [15] of the SH concept and the forms of the two major terms in SH, i.e. the zero-field splitting (ZFS) or equivalently fine structure (FS) and Zeeman electronic (Ze) Hamiltonians, provides a thorough literature survey till early 1987. An updated account of the development and present status of the SH formalisms has been provided in the review [2]. Most recently two other pertinent reviews appeared. On the basis of the available literature evidence, which implies the positive answer, the question: *"Can the EMR techniques measure the crystal (ligand) field parameters?"* has been critically examined in [16]. The distinction between, on the one hand, the actual CF/LF related quantities and, on the other hand, the actual ZFS/FS quantities, has been elucidated [16]. The intricacies of the SH theory underpinning the experimental EMR studies of paramagnetic species with the spin S≥1, especially transition ions, have been summarized in [17]. Examples of the intricacies awaiting unwary spectroscopists, drawn from recent EMR literature, have been discussed in order to illustrate the potential pitfalls and their consequences [17].

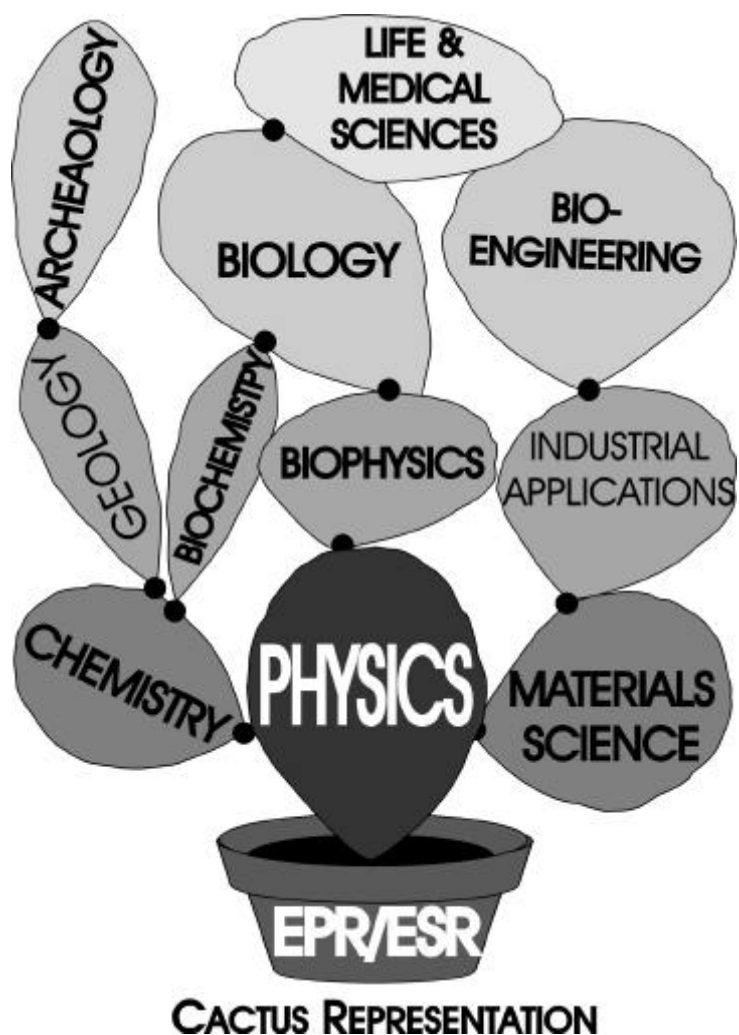

Fig. 1. The development of EPR/ESR into an EMR cactus.

The overall situation in the EMR area may be succinctly summarized as follows. At present there exists, on the one hand, a variety of the experimental EMR techniques available at the disposal of researchers, whereas on the other hand, an abundance of often inconsistent and conflicting notations for the operators and ZFS parameters used for the same physical system. The first aspect is definitely a positive development indicating the strength and versatility of the EMR techniques. However, the second one, most evident in the EMR of transition ions in crystals, is a worrying factor. The abundance of notations has its roots in the historical development of the theoretical framework underlying EMR. As our understanding progressed, it has turned out that some notations had serious deficiencies, which limited their applications to higher spin systems and lower symmetry cases. As identified in [2, 15-17], the abundance of notations together with an apparent lack of clear definitions of the basic theoretical concepts have, unfortunately, contributed to appearance and subsequent proliferation of a number of terminological misconceptions. More importantly, several incorrect relations for the SH parameters have appeared as discussed in [17-19]. The current situation has become so acute that further proliferation of these problems may hinder future healthy development of the EMR area.

### 3. Rationales for Unification of Spin Hamiltonian Notations

In this Section we consider the question why we unification of SH notations used in the EMR area is needed. First, we provide a brief background. Since the EMR area is by now mature, not many theoretical physicists or chemists work on its foundations these days, unlike in the hey days of 50's and 60', when the development of then EPR, later known also as ESR, was most rapid. Judging by the increasing occurrences of the problems outlined below and in Section 4, it appears that the understanding of some theoretical aspects of EMR among researchers has, in general, been fading. The non-specialized international EMR-related conferences, e.g. the Congress Ampere meetings [20] and APES symposia [4-6], gather a mixed group of researches reflecting the various areas of applications of the EMR techniques. While the 'language' of experimental EMR facilitates communication (so further improvements in this direction are proposed in Section 5), there is a great demand for working out an acceptable equivalent of Esperanto language to communicate more clearly the theoretical aspects of EMR.

Interestingly, in our group of budding theoretical physicists, R. Micnas, now an expert on superconductivity (see, e.g. [21]) remarked in the early 70's that the only (scientific) 'pleasure' left in the crystal field theory area (including SH theory) is to 'change notations'. The actual developments reviewed in [2, 15-19] show, regrettably, how pertinent that remark was. This, by no means, undermines the tremendous achievements of the optical and EMR spectroscopy evidenced by their wide experimental applications. Historical reasons contributing to such developments have been discussed in [2, 15]. The technique of the operator equivalents (see, e.g. [22]), which resulted in the introduction of the (*conventional*) Stevens operators (see, e.g. [8, 9]), is just one example of applications of the Wigner-Eckart theorem (see, e.g. [23]). Although it was not realized so at the time of their invention, in fact, all operator notations used nowadays in the EMR and crystal field area are equivalent to each other in one way or another by the virtue of this theorem. With the spread of EPR/ESR into other fields outside physics, the grasp of the theoretical foundations among EMR practitioners has naturally been decreasing. This has lead to various terminological and conceptual misunderstandings as well as, in most severe cases, incorrect relations creeping into the EMR literature as reviewed in [2, 15-19].

The results of the up-to-now efforts aimed at working out guidelines for unification of SH notations and data presentations in EMR have been rather limited. Kon [24] prepared a report of the Physical Chemistry Division Commission on Molecular Structure and Spectroscopy detailing the recommendations for EPR/ESR nomenclature and conventions for presenting experimental data in publications. These so far only published guidelines [24] have been limited to paramagnetic species with the spin $S=1/2$. Due to the discussions at the 31[th] Congress Ampere the author has learnt about the paper on conventions for tensor quantities used in NMR, NQR and ESR [25]. Interestingly, the empirical ideas [25] turn out to be implicitly related to the idea of orthorhombic standardization of ZFS Hamiltonians [26] generalized to the various tensors used in spectroscopy, e.g. the **g**-tensor [27]. Our recent SCI search for the keywords: nomenclature, conventions, unification, notation, and symbols, in combination with the keywords: EMR, or EPR, or ESR, has indicated but two pertinent papers, i.e. [24, 25].

The lack of any guidelines for data presentations, especially conventions for the axis systems, as well as the confusing abundance of existing notations for operators and ZFS parameters for paramagnetic species with the spin S≥1 makes it increasingly difficult to handle in a uniform way the available EMR data, especially for higher spin systems and low symmetry cases (see, e.g. [28, 29]).  The author' s ongoing efforts to improve the situation in the EMR area in this regard have been both general and specific. The general efforts include a proposal for standardization of symbols in EPR [30] and a survey of the EPR community on the EPR database [31] as well as the critical reviews on: the concept of spin Hamiltonian and notations used in EPR [2, 15], the relationships between the ZFS parameters and the CF/LF ones [16], and major intricacies concerning EMR of the spin S≥1 systems [17]. The specific efforts include development of the useful computer programs and papers dealing with specific applications. The existence in the literature of the multitude of ZFS notations and the incompatible orthorhombic and lower symmetry ZFS parameter sets as well as the usage of different axis systems to express the SH parameters [2, 15] necessitates performing various conversions and standardization calculations as well as transformations of SH parameter sets. To facilitate the most frequently needed manipulations of EMR data, a multipurpose computer package CST for conversions, standardization and transformations of SH and CF Hamiltonians [32, 33] has been developed. The package CST together with the Manual [34] is available free from the author upon request. The applications of the program CST to several cases requiring the orthorhombic and monoclinic standardization of SH parameters for transition ions in crystals have been dealt with in [29, 35, 36]. A computer program COMPARE for comparative analysis of EPR data for low symmetry paramagnetic centers has also been developed [37] and applie d in [28].

The **most compelling rationale** for unification of SH notations is as follows. For triclinic symmetry the number of the ZFS parameters of the rank: k = 2, 4, and 6 is equal to 5, 9, and 13, respectively. This translates into 5, 14, and 27 ZFS parameters in total required for, e.g. to the ion (spin) systems like: $Cr^{3+}$(3/2), $Fe^{3+}$(5/2), and $Gd^{3+}$(7/2), respectively. At the present count, there exist in the literature altogether over **twenty** different ZFS parameter notations [2, 15].  Various types of the conventional notation as well as tensor operator notations, including several tesseral-tensor operator notations and spherical tensor operators notations have been reviewed in details in [15].  A brief general and updated classification of the major operator and ZFS parameter notations used in the literature is provided in [2].  The extent of the usage of these notations in the literature have been surveyed in [30] and categorized into three groups: major, occasional, and ephemeral usage. An update on the SH operator and parameter notations used in EMR and related spectroscopies is planned in the near future. Since one must deal with the large number of the ZFS parameters for low symmetry cases and high spin S values, the confusion arising from the existence of several notations magnifies in proportion. Note also that the same operator as well as parameter symbols encountered in the up-to now literature often mean different entities.

The data accumulated so far [2, 15-17, 30, 31] may serve as a background for a meaningful decision of the EMR community with regards to selection of the reference notation(s) to be uniformly accepted. It should be emphasized that by no means such unification of notations would stifle the scientific progress in this area, as some might have feared. On the contrary, such option, if widely implemented, would release more researchers' time that at present may be lost on figuring out correlations between different notations or ensuring the actual reliability of the ZFS parameters. Adopting the two reference notations as proposed in Section 6 would reduce the multitude of notations for operators and ZFS parameters to a manageable size suitable for implementation in a computer database (see Section 5.2)
.

**Other rationales** for unification of notations stem from the need for rectification of the problems, which, for the sake of clarity, may be categorized as follows.

*(1) Incorrect relations for ZFS parameters* - This includes the following cases discussed in [18] (where source references are provided): (i) the relation between the rhombic ZFS parameters - $E$ (conventional) and $B_2^2$ (extended Stevens), which leads the maximum rhombic limit differing from the standard one defined in [26] as well as renders the tabulation of transition ion data in a major EPR/ESR handbook unreliable; (ii) the relations between the triclinic ZFS parameters - $D_{ij}$ (conventional) and $b_k^q$ (extended Stevens), which are intrinsically incorrect and thus yield wrong conversions; and (iii) the omission of the scaling factors $f_k$ in $H_{ZFS}$ = $\sum_{kq} f_k b_k^q O_k^q$ ($S_x$, $S_y$, $S_z$), which introduces an ambiguity in the meaning of ZFS parameters: $b_k^q$ or $B_k^q$ (see Section 6).

*(2) Incompatible orthorhombic and lower symmetry ZFS parameter sets* - This includes the following three cases. (i) The incompatible ZFS parameter sets for orthorhombic symmetry expressed in various non-standard systems, i.e. with the rhombicity ratio out of the standard range defined as: $0 \leq \lambda' = B_2^2 / B_2^0 = b_2^2 / b_2^0 \leq 1$ [26]. The structural implications of the non-standard $\lambda'$ pointed out in [26, 29, 35, 36] indicate clearly that standardization is necessary for proper data interpretation, since a very large 'rhombicity' measured by some authors appears to be an artifact only. (ii) The application of mixed axis systems for the same ion/host case, i.e. a *cubic* system for the 4[th]-order ZFS terms, whereas a *rhombic* system for the 2[nd]-order ones. This necessitated additional conversions for proper data analysis and comparison. (iii) Lack of proper understanding of the symmetry properties of monoclinic SH discussed in [38-40]. These properties allow three distinct yet equivalent SH forms for each choice of the axis system with respect to the monoclinic $C_2$ axis, namely, $C_2 \| Z$, $C_2 \| Y$, or $C_2 \| X$, each yielding different yet equivalent ZFS parameter sets. In many cases no clear definition of the axis system is provided rendering the data at best undefined or at worse based on an inappropriate form of ZFS [29] (as well as CF [41]) Hamiltonian.

The above considerations amply exemplify the rationales for unification of notations and adoption of standard axis systems for presentation of ZFS parameters for the spin S≥1 systems. Keeping in mind that the current situation hampers development of the EMR database proposed in Section 5, adequate measures to improve the notation aspects are indispensable. Various ways in this regard, which require a concerted effort within the EMR community at the international level, are suggested in Section 6.

## 4. Rationales for Unification of EMR Nomenclature

In this Section we consider the question why unification of nomenclature used in the EMR area is needed. The collective attempts in similar direction have a long tradition in other areas of spectroscopy. For example, under the auspices of appropriate IUPAC Commissions groups of researchers have managed to work out, e.g. *"Glossary of Terms used in Photochemistry"* [42] and the document *"Names, Symbols, Definitions and Units of Quantities in Optical Spectroscopy"* [43]. The motivation for such documents has been clearly stated in [42]: *"The purpose of the Glossary is to provide definitions of terms and symbols commonly used in the field in order to achieve consensus on the adoption of some definitions and on the abandonment of inadequate terms"*. Collective efforts to work out a glossary of terms in the EMR area are at the beginning. Apart from the first and so far only attempt in this direction lead by Kon [24], no equivalent of, e.g. the documents [42, 43], exist in the EMR literature. It does not mean that there are no problems with ambiguity or uncertainty in the usage of some EMR terms. On the contrary, the situation is rather messy as discussed below.

**Major conceptual problems** concerning some EMR terms may be classified into two general groups as follows.
*(1) Distinction between various SH formalisms:*
The reviews [2, 15] of the microscopic SH (MSH), generalized SH (GSH), and phenomenological SH (PSH) have indicated a wide spread of serious misconceptions concerning the various SH formalisms in the EMR literature. The physical origin of the SH formalisms appears not to be well understood by several authors leading to serious problems, which need clarifications [2, 15-19]. A survey of the conceptual confusion between various major EMR notions has also been provided in [2, 15]. It appears that meanings of several crucial terms are often confused with each other, e.g. physical *versus* effective Hamiltonian, real *versus* effective *versus* fictitious spin, and generalized SH *versus* phenomenological SH.
*(2) Confusion between the ZFS Hamiltonian and other physically distinct Hamiltonians:*
The physical nature of the zero-field-splitting (ZFS) Hamiltonian as well as parameters as presented by several authors is often incorrect and needs clarifications. This includes the relationship between the ZFS Hamiltonian and (a) the crystal-field (CF), (b) electronic spin-spin, and (c) nuclear-quadrupole Hamiltonians [15]. The origin and possible roots of the incorrect terminology consisting in mixing up the physically distinct quantities at different levels have been examined in [15]. While the instances of the confusion of the type (c) were only sporadic [15], the other two types (a) and (b) are widely spread and have serious consequences, much beyond semantic inconsistencies. The confusion of the type (b) has lead to misinterpretation of the relationships between $B_k^q$(ZFS) and the various *actual* spin-spin parameters, consequences of which have been discussed in [2, 15]. A recent survey of literature has revealed that an update on [2, 15] concerning this confusion in EMR and, moreover, in magnetism would be timely. The most widely spread confusion is that of the type (a), i.e. between the ZFS and CF quantities [2, 15, 16]. The most serious consequences include: (i) using the specific ZFS parameter notation for the actual CF parameters or *vice versa,* (ii) applying incorrectly the point-charge model for ZFS parameters, and (iii) creating an impression that the EMR techniques can measure the crystal (ligand) field parameters. A large number of cases of incorrect terminology concerning this confusion and other inconsistencies identified in recent literature have been systematically classified in [16]. Implications of the confusion in question are serious indeed. The incorrect terminology has lead to misinterpretation of the experimental EMR data as well as contributed to misleading keyword classifications of papers in journals and as a consequence reduced the reliability of the retrieval of references from scientific literature databases [16].

**Other specific conceptual problems** as well as, to a certain extent, notational problems may be classified as follows.

*(i) Inadmissibility of the odd-order (k =1, 3, 5) ZFS terms:*

Their introduction arises from misunderstanding of the properties of GSH as discussed originally in [44]. An updated discussion in [2] indicates that the matter is not put to rest yet. Regrettably, the confusion concerning the form of the SH triggered by this problem proliferates as indicated by a recent referee report incorrectly interpreting the monoclinic SH forms [45].

*(ii) Confusion concerning derivation of MSH for $3d^5$ ions at axial symmetry:*

Here an erroneous identification of the wave-functions for the *physical* and the *effective* spin Hamiltonian has resulted in invalid microscopic SH relations for ZFS parameters as discussed in [19].

*(iii) Usage of truncated forms of ZFS Hamiltonian:*

This includes cases of (a) some independent ZFS terms set to zero, (b) omission of one of several interdependent ZFS terms, and (c) neglect of some operator terms leading to specific definition of ZFS parameters used, e.g. in high-magnetic-field and high-frequency EMR (HMF-EMR) and magnetization studies of $Mn_{12}$ (S=10) complexes [17].

The above summary of the nomenclature problems existing in the EMR literature, especially EMR of transition ions, indicates that the need for unification of nomenclature is overwhelming. Keeping in mind the current situation further proliferation of the incorrect terminology must be prevented. Such positive outcome would also lead to increased reliability of the published EMR data. However, a concerted effort within the EMR community is indispensable. Various ways in this regard at the international level are suggested in Section 6.

## 5. EMR Data as 'Community' Resources

The various categories of data generated during EMR studies, which constitute 'community resources', are discussed in a comprehensive way in this Section. A major category consists of the experimental and theoretical parameters describing EMR spectra as well as data on the experimental conditions at which these parameters were measured or the values of more microscopic parameters which enable interpretation of the SH parameters. We name collectively these datasets as 'EMR data' to distinguish them from other EMR-related data, e.g. data about the EMR facilities or bibliographical data. Hence, below we refer to a computerized database of the experimental and theoretical EMR parameters and relevant data as 'EMR database'.

In should be kept in mind that a satisfactory and widely accepted solution to the notational problems as well as, to a certain extent, the conceptual ones, outlined in Sections 3 and 4, respectively, is an indispensable prerequisite for development of an EMR database as first argued in [30]. As a preparation for development of such database a survey of the EMR community needs in this regard has been carried out [31]. More recently, the author, in the capacity of the APES President, has also presented the idea of EMR database as a discussion item at the consecutive meetings of the APES [4-6]. Although, generally the EMR researchers are in favour of the proposed idea, it turns out that due to apparent lack of resources and other pressing issues this idea may still have to wait long for implementation, unless more proactive measures proposed in Section 6 are adopted. It appears that at present there is a growing realization of the opportunities created by the advances in the IT and Internet technologies, which were unavailable in the early 90's. Hence, there is a strong rationale for utilization of the capabilities of the new technologies to provide a useful tool for EMR researchers, far beyond the existing bibliographic literature databases. A central issue to any computerized database is the selection and organization of data items to be stored. Hence, the data items pertinent for the EMR database are discussed in Section 5.1. The EMR-related databases under development at the regional scale as a service activity of the APES for the Asia-Pacific EMR researchers as well as other pertinent issues are outlined in Section 5.2.

*5.1 EMR data and their presentation – a step towards EMR database*

To facilitate the process of extracting from original sources the data relevant to our current research projects, and their subsequent meaningful analysis, three types of Data Entry Forms have recently been worked out: (1) EMR Form - for experimental and theoretical EMR data, including the Zeeman g-factors and ZFS parameters, (2) CFT Form - for CF parameters and optical spectroscopy data, and (3) MBS Form - for the ZFS parameters derived from Mössbauer spectroscopy and magnetic susceptibility studies. To speed up the data entry process, whenever possible, various predetermined options for parameter symbols and notations have been provided in a tick box format. Each Data Entry Form is accompanied by a checklist of the most intricate points that need to be paid attention to for a meaningful extraction of data from the original sources. In the case of the EMR Form, the checklist is based on the intricacies outlined in Section 3 and 4. Additional basic explanations common to the three Data Entry Forms have also been worked out. Copies of the Data Entry Forms and related documentation can be obtained from the author upon request.

Work on the current extensive reviews in the area of EMR and CF theory may be greatly facilitated by the use of EMR and CFT Data Entry Forms. An additional intended purpose of the EMR Data Entry Form has been its potential use as a major tool for systematic data processing and subsequent entry into the future EMR database. The scope of the three Data Entry Forms illustrates the interdependencies between the various fields for which the SH is a central concept. A broader approach to the SH-related data, which makes use of the correlation between physically equivalent data obtained from different experimental techniques, seems to be more useful that a narrow one. The CFT and MBS Data Entry Forms may be utilized in future extensions of the EMR database to other relevant data.

Below we have identified the major data items of importance for adequate analysis and comparison of the results of experimental and, to a certain extent, also theoretical, EMR studies. This selection is based mostly on the data comprised in the present EMR Data Entry Form supplemented with a few data items taken from the general ones included in the APES Membership Form (see Section 5.2). This listing is not fully comprehensive and is intended as a starting point for further fine-tuning of the guidelines for presentation of EMR data in consultations with the EMR community. The indentation is used below to indicate groups of related data items and their hierarchy.

- Compound = *name & formula*
  - Sample type = *Single crystal / Powder / Glass / ?* ['?' means: unspecified sample type]
    - Band / Frequency range [GHz]
      - $B_o$ max [T] & Mode: $B_o \perp B_1$ *or* $B_o \parallel B_1$
        - Temperature range [K]
- Crystallographic data = *(a, b, c) units*
  - Space group = *Int. Tables for Crystallography. Ref. No.*
    - Angles [deg] (axis1, axis2) = *Axis system w.r.t. (a, b, c)*
      - Number of magnetically inequivalent sites
        - Position label
- Ion (valence) *or* center type (spin) = *ion – ligands complex*
  - Local site symmetry [LSS] = *Point group symmetry / class*
    - CF orbital ground state *(for TM ions only)*
      - Charge compensation & vacancy
- Symmetry class / type *assumed* for SH
  - Original Form of SH used for data analysis
    - Original notation for ZFS parameters & operators
      - Original Units for ZFS parameters
        - Software used for fitting

The above suggestions are pertinent for both the spin S=1/2 and S≥1 systems and hence may be used for extension of the recommendations for EPR/ESR nomenclature and conventions for presenting experimental data in publications pertinent only for S=1/2 [24]. It appears that adoption of the unified data items and presentation guidelines for EMR data may help to (i) reduce confusion about the notations used, (ii) prevent incorrect relations and misinterpretations of data, and (iii) increase the overall quality of EMR data published in the literature. Hence further development of these guidelines and their subsequent implementation is deemed essential.

*5.2 EMR-related databases*

The major initiative in this category concerns establishment of the EMR database as defined above. The activities described in Section 5.1 provide the necessary preparatory framework. One more preparatory aspect worth mentioning is the author's literature database (LDB for short) developed in *Excel* to the present form over many years. LDB covers major physics and, to a certain extent, chemistry journals related mostly to the EMR, crystal field and optical spectroscopy, and magnetism areas, with special focus on the studies of transition ions. At present it contains over 12,000 entries on the reprints stored in the author's private collection organized in a systematic way into a system of physical files and more recently and increasingly so also of electronic files. The main purpose of LDB is to facilitate the author's research work and to handle the vast collection of reprints accumulated since 1970. Its structure has evolved with experience of the main user. Primarily, this database was designed in such a way as to minimize time used for the data entry process. Hence the bibliographical data are not fully comprehensive. The content of LDB reflects the author's changing priorities and focus interests. Hence it could be also useful for other researchers working in similar areas, especially if they need to find quickly data on particular topics (as well as ions and compounds), which are not directly searchable in the commercial literature databases. There are at present 44 specialized topics included in the LDB in the field "T/L" (Topical List). To name a few topics of interest to EMR and magnetism studies: 'D/O' – definitions of operators and SH notations discussed in Section 3, 'CF=ZFS' - the confusion between the CF and ZFS quantities discussed in Section 4, 'diso/ZFS' - the studies of disorder in EMR spectra described by ZFS Hamiltonian, 'EI&ZFS' - MSH contributions to the actual ZFS parameters from the exchange interactions (EI) as well as dipolar and/or spin-spin interactions, 'EI=ZFS' - the cases identified in the literature where the name 'exchange (or spin-spin) interactions' are confusingly used for the actual ZFS Hamiltonian, 'HOZT' - higher order Zeeman terms: e.g. $BS^3$, $B^2S^2$, $B^3S$, and 'NS/ZFS' - nonstandard ZFS parameters. The T/L tag: 'S=2' indicates cases of the spin S=2 system originating either due to a particular ion (as listed in the field: 'Ions') or an exchange coupled system (such papers are usually kept in the file coded: 'EPR/pairs'). The 'S=2' tag enables an efficient identification of the papers providing a clear indication of the orbital singlet ground state for the $3d^4$ and $3d^6$ ions. Since for most of these ions the ZFS is rather large or very large, these cases are suitable for HMF-EMR measurements. By searching the fields: 'IONS' and/or 'COMPOUNDS' one can identify the cases of the $3d^4$ and $3d^6$ ions and materials. Most recently these aspects of LDB have been utilized for the review of magnetically ordered high-spin S=2 $Fe^{2+}$ ion systems potentially suitable for HMF-EMR studies [46]. The database LDB provides other important information and enables searches for various parameters, e.g. the most recent years of publication, papers on specific compounds or ions, etc, using the *"Filter"* options within *Excel*. The data and references retrieved from LDB may also serve a good starting point for more specialized searches of commercial databases. LDB has recently been provided free of charge to a number of researchers and proved to be useful, especially for EMR studies. Potentially LDB can be used as a source of primary data and a prototype for a full scale computerized database of the experimental and theoretical EMR parameters and relevant data.

Other EMR-related databases initiatives coordinated at present by the author in the capacity of the APES President include (i) the APES Membership database and (ii) the Directory of EPR/ESR Facilities within the Asia-Pacific Region, which lists the useful information, e.g. available spectrometers, major research interests, and projects, etc. It may be envisaged that coordination of these databases with those independently maintained by other EMR-related organizations, especially the International EPR/ESR Society (IES), would be beneficial to all organizations as well as the whole EMR community. Recently, the Presidents of the IES (Prof. J. Pilbrow) and APES (the author) has discussed a possible merge of the two societies Membership databases into one combined platform. This is a formidable task, which can only be achieved under the framework outlined in Section 6.

### 6. Blueprint for Future Actions

In this Section, first we outline a set of tentative guidelines arising from the rationales presented in Section 3 and 4. Next, we consider ways of achieving a consensus on the notations and nomenclature standards in EMR as well as on the guidelines for EMR data presentation outlined in Section 5.1. Finally, the framework for establishment of EMR-related databases outlined in Section 5.2 is provided. The *'unity in diversity' concept*, which is very valuable for, e.g. biodiversity, is definitely not appropriate for scientific notation and nomenclature as argued in Section 3 and 4. On the contrary, we need standardization of notations and nomenclature as well as guidelines for data presentation in order to improve the quality of EMR data and increase the overall reliability of data taken from various sources.

Adopting the two notations, defined below, as the reference notations would greatly improve the overall situation in the EMR area in future. Moreover, it would also reduce the multitude of notations for operators and ZFS parameters to a manageable size suitable for implementation in a computerized EMR database discussed in Section 5.2.

*Major reference notation:*
**The extended Stevens** (ES) operators defined in [47]:

$$H_{ZFS} = \sum_{kq} B_k^q O_k^q (S_x, S_y, S_z) = \sum_{kq} f_k b_k^q O_k^q (S_x, S_y, S_z). \quad (1)$$

This notation can be used for the 2, 4, and 6-order ZFS terms and after proper extension also to the higher-order ZFS terms required, e.g. for the spin S>7/2 systems. Note that the transformation properties of the ES operators and parameters have been worked out [47, 48]. The parameters $b_k^q$ with the consistent definition of the factors [2, 15]: $f_2 = 1/3$, $f_4 = 1/60$, $f_6 = 1/1260$, would be preferable. Adoption of the parameters $B_k^q$ may exacerbate the confusion discussed in Section 4, since this symbol has also been often used for the CF parameters: $H_{CF} = \sum_{kq} B_k^q O_k^q$ (**L** or **J**) in the literature [2, 15-17].

*Supplemental reference notation:*

**The conventional notation** defined in [15] (see also [8-10], with the correction to [9] given in [18]):

$$H_{ZFS} = \mathbf{S \cdot D \cdot S} = D[S_z^2 - \tfrac{1}{3}S(S+1)] + E[S_x^2 - S_y^2]. \tag{2}$$

This notation shall be allowed only for the second-order ZFS terms [15] and for axial symmetry (*E*=0) as well as orthorhombic and monoclinic or triclinic symmetry in the principal axis system. Note that several problems inherent for this type of notations as well as the inconsistencies existing in the literature summarized in [15] (see also [16, 18]) speak against using the conventional SH notations for a wider purpose. However, in view of its popularity among EMR researchers (see, e.g. [49-51]), especially those working on chemical, biological, and materials science applications of EMR to the spin S ≤ 1 systems, usage of the conventional SH notation defined in (2) as a **supplemental** reference notation seems advisable.

Unification of EMR data presentation requires also decision on two other aspects. Concerning the units, it is proposed to adopt as the s*tandard* units [$10^{-4}$cm$^{-1}$] *or* [cm$^{-1}$] for the ZFS parameters $b_k^q$ [15]. Concerning the axis systems, the standard range defined in [26]: $0 \leq \lambda' = B_2^2 / B_2^0 = b_2^2 / b_2^0 \leq 1$ be best adopted for orthorhombic and lower symmetry cases. Note that the two options have been implemented as default options in the CST package [33, 34].

The above proposals are based on a thorough analysis of the prevailing trends in the EMR literature [2, 15, 16] and correspond well with the notations actually used in the major EMR textbooks [8-10, 49-51]. Support for the above proposal comes from a number of researchers consulted directly by the author as well as indirectly from Prof. K.W.H. Stevens, who stated in his recent book: '*There is a problem over listing the replacement angular momentum operators and their matrix elements because of the diversity of the definitions and errors in the tabulated matrix elements. The reader is therefore advised to consult the papers by C. Rudowicz, particularly,* [Refs 47, 48, 15 - below] *for a critical account of the current literature and for proposals for future standardization in the definitions and notations. A step forward would seem to be to adopt his suggestions*'.

Judging by the SCI searches indicating an increasing number of citations, especially of the review [15] and the paper [47], the proposed notation standards as well as the guidelines for EMR data presentation for transition ions at orthorhombic and lower symmetry [26, 29, 33-35] have received substantial recognition in the last several years. Adoption of a set of widely accepted nomenclature standards in EMR, at the level comparable to that in other areas, e.g. NMR and Mössbauer spectroscopy, requires concerted efforts of the international EMR community described below.

There may be some initial resistance to use a particular notation from various researchers who have developed their own fitting and simulation computer programs incorporating specific notations other than the standard ones proposed above. However, it should be kept in mind that the aim of unification of notations and unification of nomenclature is not narrowly defined 'uniformity at all costs' but (i) improving the quality of EMR research and facilitating EMR researchers work as well as (ii) getting rid, in the long term, of the problems identified in Section 3 and 4. These problems have been succinctly summarized by an anonymous referee of [18]: *"Poorly defined notation and lack of any universally accepted definitions of spin-Hamiltonian forms has been a long standing confusing feature of EPR"*. The current situation calls for adoption, as an urgent imperative, of unified guidelines for presentation of the ZFS parameters in the EMR-related literature. It may be hoped that the widely accepted standards of EMR data presentation will improve in the long run the general understanding of the intricate aspects of EMR among experimentalists and theorists alike.

The major way of achieving a consensus on the notations and nomenclature standards in EMR as well as on the guidelines for EMR data presentation is to convince as many EMR researchers as possible about the perceived benefits of these proposals. The unification attempts in the NMR area [25] show the need for coordination of individual and group efforts as well as proper support from international bodies and societies. Judging by the experience of the researchers in other areas [25, 42, 43], implementation of the proposed unification of notations and unification of nomenclature in the EMR area can only succeed if not left to individual attempts. The role of the author and APES in spearheading these efforts will not be sufficient in the long run. Various forms of support from the EMR community worldwide and the pertinent international organisations, e.g. IES, Groupement AMPERE, ISMAR, and UPAC, are indispensable. To enable proper channelling and utilization of this support an adequate organisational framework is needed as outlined below.

Each proposal outlined above requires some financial resources as well as efforts coordinated at the international level. These requirements vary for various activities - drafting a Glossary of Terms used in EMR requires least funds but much coordination of efforts, whereas the establishment of a computerized EMR database will be the most costly and labor intensive endeavor. All the activities necessary for implementation of these proposals could be best carried out if their coordination is combined under one roof by creation of a dedicated **International EMR Support Unit**. This organizational framework would enable to achieve maximum benefits in a relatively short time span and in the most economical way. On the other hand, practical implementation of particular activities may be carried out as appropriate, i.e. either within the Unit or in other EMR centers all over the world. By taking full advantage of the present capabilities of the INTERNET to coordinate scientific efforts, the most efficient sharing of workload may be achieved with minimal funds. Such *'distributed yet centrally coordinated network'* seems the best organisational framework. This option requires finding a **host institution** for the proposed Unit. Preliminary talks held by the author in summer 2002 give grounds for cautious optimism in this regard. A search on a wider scale for a host institution may start after a proper agreement between IES and APES is reached. In order to convince potential institutions to host the proposed Unit a well design plan and clear objectives must be presented. Adequate benefits for the host institutions shall be built into the proposal to make hosting the Unit an attractive option. The major benefits arising from hosting the Unit in the long terms would be the enhancement of local EMR-related research and strengthening of the interactions with international EMR groups, which shall result in a greater access to well-established facilities, as well as improved funding and literature resources. Below the proposed objectives, focus activities, and structure of the International EMR Support Unit, which satisfy these criteria, are outlined.

The following **objectives** arising from the needs of the EMR community outlined above are envisaged.
   (a) To carry out **research in the foundations of EMR** aimed at providing better tools for understanding and interpretation of experimental results;
   (b) To create and maintain **EMR-related computer** databases, which would facilitate EMR-related research;
   (c) To provide **support for EMR international organisations**, particularly IES and APES, under the auspices of which the Unit is to operate;
   (d) To **participate in organization** of the EMR-related events and meetings aimed at dissemination of scientific information;
   (e) To **foster other EMR-related activities** in collaboration with international organizations, which would lead to a better utilization of resources.

The following **focus activities** arising from the proposals outlined above are envisaged.
- Coordination of the work on the unification of notations used in EMR and the recommendations for EMR nomenclature and conventions for the spin $S>1/2$ systems.
- Work on a draft Glossary of EMR Terms and coordination of further consultations within the EMR community.
- Development and/or coordination as well as standardization of comprehensive computer packages and programs facilitating EMR research.
- Organization and maintenance of computerized EMR databases.
- Ongoing research work on topical reviews as well as expert reports covering various specialized areas of EMR.
- Work on the standards for EMR measurements in order to make the EMR techniques more widely accessible and applicable, especially aiming at enhancement of the industrial applications of EMR.
- Support of initiatives and activities leading to the development of the state-of-the-art EMR facilities at the regional level with respect to the host institution.

Concerning the **structure** of the International EMR Support Unit – the Unit is meant to be an international and inter-disciplinary as well as, at a later stage, an inter-institutional (at the regional level) research unit. It is intended to operate under the auspices of the International EPR/ESR Society and the Asia-Pacific EPR/ESR Society affiliated with IES. At a later stage, after proper agreements are reached, extension to other relevant international and national organizations may be considered. A starting point is to form a small **Preparatory Group** for the establishment of the International EMR Support Unit, which in due time, in an extended form, may evolve into the Scientific Board of the Unit. The Preparatory Group may be nominated by IES and APES and mandated first to carry out a feasibility study of the establishment of the Unit, including coordination of the search for a host institution. The initial process requires action in both directions: the bottom-up approach, i.e. activities initiated by concerned researchers, as well as the top-down approach, i.e. activities initiated by IES and APES.

It is envisaged that the activities and accordingly the structure of the Unit will be developed in stages. The Unit's activities would constitute a niche area in the knowledge organization as well as the quality assurance of EMR data. For example, adoption of unified guidelines is only possible after the wide consultation within all sectors of the EMR community representing various areas of the applications of EMR techniques. The consultation process shall also help working out a Glossary of EMR terms. A draft of such Glossary could be prepared by a small group of researchers on behalf of the EMR organizations for further circulation under the auspices of an appropriate IUPAC Commission. Hence, it is proposed to form, simultaneously with the Preparatory Group for the Unit, an **International Committee for EMR Nomenclature** mandated by the IES and APES to carry out the relevant task. The International EMR Support Unit would then provide a technical support for the activities of this Committee in discharging its duties. After the initial period of consolidation, as a preparation for development of the EMR database, the Unit in this regard may carry out an updated survey of the EMR community needs.

Comments from all sectors of the EMR community on the proposals put forward in this paper are welcome. Your feedback is essential for fine-tuning these proposals. Finally, a short concluding remark will suffice: *united we shall overcome*.

**ACKNOWLEDGMENTS**

This work was partially supported by the RGC and the City University of Hong Kong grant: SRG 7001277. Special thanks are due to Prof. Stefan Jurga and Prof. Jan Stankowski for their invitation to the 31$^{st}$ Congress AMPERE and useful comments on the proposals presented at the Congress. Valuable feedback on the proposals during the Congress from Professors D. Goldfarb, S.S. Eaton, G.R. Eaton, J. Pilbrow, Kev Salikhov, and D. Stehlik, as well as S.K. Hoffmann, R. Micnas, and N. Pislewski is also gratefully acknowledged. Initial discussions with Professors K. Dyrek, L. Proniewicz, and Z. Sojka have greatly contributed to the proposals – the authors is very grateful for their overall support. Thanks are due to Dr E. Roduner for bringing reference 25 to our attention. The author would like to thank all coworkers and co-authors, especially Dr A.A. Galeev, Prof. S.K. Misra, and Miss H.W.F. Sung. Technical assistance from Miss H.W.F. Sung in preparing this paper is gratefully acknowledged.